\documentstyle[sprocl,epsfig]{article}

\newcommand{\li}{\mathop{{\mbox{Li}}_4}\nolimits}

\begin{document}

\title{\vskip-2.cm{\baselineskip14pt
\centerline{\normalsize\hfill MPI/PhT/98--87}
\centerline{\normalsize\hfill hep--ph/9811466}
\centerline{\normalsize\hfill November 1998}
}
\vskip.5cm
DECOUPLING RELATIONS, EFFECTIVE LAGRANGIANS AND LOW-ENERGY
THEOREMS\footnote{To appear in the
{\it Proceedings of the IVth International Symposium
on Radiative Corrections (RADCOR~98)},
Barcelona, Spain, 8--12 September 1998, edited by J. Sol\`a.}
}

\author{BERND A. KNIEHL}

\address{Max-Planck-Institut f\"ur Physik (Werner-Heisenberg-Institut),\\
F\"ohringer Ring 6, 80805 Munich, Germany\\
E-mail: kniehl@mppmu.mpg.de}

\maketitle\abstracts{If QCD is renormalized by minimal subtraction (MS),
at higher orders, the strong coupling constant $\alpha_s$ and the quark
masses $m_q$ exhibit discontinuities at the flavour thresholds, which are
controlled by so-called decoupling constants, $\zeta_g$ and $\zeta_m$,
respectively.
Adopting the modified MS ($\overline{\mbox{MS}}$) scheme, we derive simple
formulae which reduce the calculation of $\zeta_g$ and $\zeta_m$ to the
solution of vacuum integrals.
This allows us to evaluate $\zeta_g$ and $\zeta_m$ through three loops.
We also establish low-energy theorems, valid to all orders, which relate the
effective couplings of the Higgs boson to gluons and light quarks, due to the
virtual presence of a heavy quark $h$, to the logarithmic derivatives w.r.t.\
$m_h$ of $\zeta_g$ and $\zeta_m$, respectively.
We also consider the effective QCD interaction of a CP-odd Higgs boson and 
verify the Adler-Bardeen nonrenormalization theorem at three loops.}

\section{Introduction}

It is generally believed that quantum chromodynamics (QCD) is the true theory 
of the strong interactions.
There are still open questions, concerning the origin of confinement or as to 
why the quark masses $m_q$ and the asymptotic scale parameter $\Lambda$ have 
the values they happen to have.
The answers to these questions probably lie outside the scope of perturbative 
QCD, which forms the basis of this presentation.
In perturbative QCD, the strong coupling constant $\alpha_s=g^2/(4\pi)$, where
$g$ is the gauge coupling, is small enough to serve as a useful expansion
parameter, and quarks and gluons may appear as asymptotic states of the
scattering matrix.

QCD is a nonabelian Yang-Mills theory based on the gauge group SU(3).
In the covariant gauge, the Lagrangian reads
\begin{eqnarray}
{\cal L}&=&\sum_{q=1}^{n_f}
\bar\psi_q^i\left(i\;\slash\!\!\!\!D^{ij}-\delta^{ij}m_q\right)\psi_q^j
-\frac{1}{4}\left(G_{\mu\nu}^a\right)^2
-\frac{1}{2\xi}\left(\partial^\mu G_\mu^a\right)^2
+\left(\partial^\mu\bar c^a\right)\nabla_\mu^{ab}c^b,
\nonumber\\
D_\mu^{ij}&=&\delta^{ij}\partial_\mu-ig[T^a]^{ij}G_\mu^a,\qquad
\nabla_\mu^{ab}=\delta^{ab}\partial_\mu-gf^{abc}G_\mu^c,
\nonumber\\
G_{\mu\nu}^a&=&\partial_\mu G_\nu^a-\partial_\nu G_\mu^a
+gf^{abc}G_\mu^bG_\nu^c,
\label{eq:qcd}
\end{eqnarray}
with $n_f=6$ flavours of quarks $\psi_q^i$ ($i=1,2,3$),
gluons $G_\mu^a$ ($a=1,\ldots,8$),
Faddeev-Popov ghosts $c^a$, and
gauge parameter $\xi$.
The generators $T^a$ satisfy the commutation relations 
$[T^a,T^b]=if^{abc}T^c$, where $f^{abc}$ are the structure constants.

In the calculation of QCD quantum corrections, one generally encounters, among 
other things, ultraviolet (UV) divergences, which must be regularized and
removed by renormalization.
In quantum electrodynamics, it is natural to employ the on-shell 
renormalization scheme, where the fine-structure constant is renormalized in
the limit of the photon being on its mass shell.
Due to confinement, this limit cannot be taken in QCD, and it is natural to 
employ the most convenient renormalization scheme instead.
It has become customary to use dimensional regularization\cite{bol} in 
connection with minimal subtraction (MS).\cite{tho}
I.e., the integrations over the loop momenta are performed in $D=4-2\epsilon$
space-time dimensions, introducing a 't~Hooft mass scale $\mu$ to keep the
(renormalized) coupling constant dimensionless.
The poles in $\epsilon$ that emerge as UV divergences in the physical limit
$\epsilon\to0$ are then combined with the bare (UV-divergent) parameters and
fields in Eq.~(\ref{eq:qcd}) so as to render them renormalized (UV finite).
This is always possible because QCD is a renormalizable theory.
In the modified MS ($\overline{\rm MS}$) scheme,\cite{bar} the specific 
combination of transcendental numbers that always appears along with the poles 
in $\epsilon$ is also subtracted.
In the following, bare quantities will be denoted by the superscript `0.'
Specifically, we have
\begin{eqnarray}
g^0&=&\mu^\epsilon Z_gg(\mu),\qquad
m_q^0=Z_mm_q(\mu),\qquad
\xi^0-1=Z_3(\xi(\mu)-1),
\nonumber\\
\psi_q^{0,i}&=&\sqrt{Z_2}\psi_q^i(\mu),\qquad
G_\mu^{0,a}=\sqrt{Z_3}G_\mu^a(\mu),\qquad
c^{0,a}=\sqrt{\tilde{Z}_3}c^a(\mu).
\end{eqnarray}
In the MS-like schemes, the renormalization constants $Z$ may be written in
the simple form
\begin{equation}
Z=1+\sum_{i=1}^\infty\sum_{j=1}^iZ_{ij}\frac{a^i}{\epsilon^j},
\end{equation}
where $a=\alpha_s/\pi$ is the renormalized couplant and $Z_{ij}$ are numbers.
I.e., the $Z$ factors do not explicitly depend on dimensionful parameters.
In particular, $Z_m$ is generic for all $q$.
A crucial advantage of the MS-like schemes is that $Z_g$ and $Z_m$ are $\xi$ 
independent to all orders.
This property carries over to $\alpha_s$ and $m_q$, so that it makes sense to
extract these parameters from experimental data.

$Z_g$ and $Z_m$ carry the full information on how $\alpha_s$ and $m_q$ run 
with $\mu$.
In fact, from the $\mu$ independence of $g^0$ and $m_q^0$ it follows that
\begin{eqnarray}
\beta(a)&\equiv&\frac{da}{d\ln\mu^2}
=-a\left(\frac{d\ln Z_g^2}{d\ln\mu^2}+\epsilon\right)
=-\sum_{n=0}^\infty\beta_na^{n+2},
\nonumber\\
\gamma_m(a)&\equiv&\frac{d\ln m_q}{d\ln\mu^2}
=-\frac{d\ln Z_m}{d\ln\mu^2}
=-\sum_{n=0}^\infty\gamma_na^{n+1}.
\label{eq:rge}
\end{eqnarray}
The Callan-Symanzik $\beta$ function and the quark mass anomalous dimension
$\gamma_m$ are universal in the MS-like schemes.
Moreover, $\beta_0$ and $\beta_1$ are universal in the larger class of schemes
which have mass-independent $\beta$ functions.
In the MS-like schemes, the coefficients $\beta_n$ and $\gamma_n$ are known
through four loops, i.e., $n=3$.\cite{ver}

To summarize, the MS-like schemes offer several advantages.
They are easy to implement in symbolic manipulation programs and are tractable
at high numbers of loops.
Furthermore, $\alpha_s$ and $m_q$ are $\xi$ independent to all orders and may
thus be regarded as physical observables.
The price to pay is that heavy quarks do not automatically decouple.
However, as will become clear in the following, the theoretical ambiguity 
associated with the matching at the flavour thresholds is negligible if higher
orders are taken into account.

\section{Decoupling of Heavy Quarks}

The decoupling theorem states that the infrared structure of unbroken, 
nonabelian gauge theories is not affected by the presence of heavy fields 
coupled to the massless gauge fields.\cite{sym}
As a consequence, a heavy quark $h$ decouples from physical observables 
measured at energy scales $\mu\ll m_h$ up to terms of ${\cal O}(\mu/m_h)$.
However, the proof of this theorem relies on the use of mass-dependent $\beta$ 
functions.
Thus, this theorem does not automatically hold for the parameters and fields
in MS-like schemes.
The standard way out is to implement explicit decoupling by using the language
of effective field theory.

As an idealized situation, consider full QCD with $n_l=n_f-1$ light quarks 
$q$, with $m_q\ll\mu$, plus one heavy quark $h$, with $m_h\gg\mu$.
The idea is to construct an effective theory, QCD$^\prime$, by integrating out
the $h$ quark.
The parameters and fields of the effective theory, which will be denoted by a
prime, are related to their counterparts of the full theory by the decoupling 
relations,
\begin{eqnarray}
g^{0\prime}&=&\zeta_g^0g^0,\qquad
m_q^{0\prime}=\zeta_m^0m_q^0,\qquad
\xi^{0\prime}-1=\zeta_3^0(\xi^0-1),
\nonumber\\
\psi_q^{0\prime,i}&=&(\zeta_2^0)^{1/2}\psi_q^{0,i},\qquad
G_\mu^{0\prime,a}=(\zeta_3^0)^{1/2}G_\mu^{0,a},\qquad
c^{0\prime,a}=(\tilde\zeta_3^0)^{1/2}c^{0,a}.
\label{eq:dec}
\end{eqnarray}
By gauge invariance, the most general form of the effective Lagrangian
${\cal L}^\prime$ emerges from Eq.~(\ref{eq:qcd}) by only retaining the light
degrees of freedom and reads
\begin{equation}
{\cal L}^\prime\left(g_s^0,m_q^0,\xi^0;\psi_q^{0,i},G_\mu^{0,a},c^{0,a};
\zeta^0\right)
={\cal L}\left(g_s^{0\prime},m_q^{0\prime},\xi^{0\prime};
\psi_q^{0\prime,i},G_\mu^{0\prime,a},c^{0\prime,a}\right),
\end{equation}
where $\zeta^0$ collectively denotes all decoupling constants of
Eq.~(\ref{eq:dec}).
The latter may be derived by imposing the condition that the results for
$n$-particle Green functions of light fields in both theories should agree up
to terms of ${\cal O}(\mu/m_h)$.

\begin{figure}
\centering
\mbox{\epsfig{file=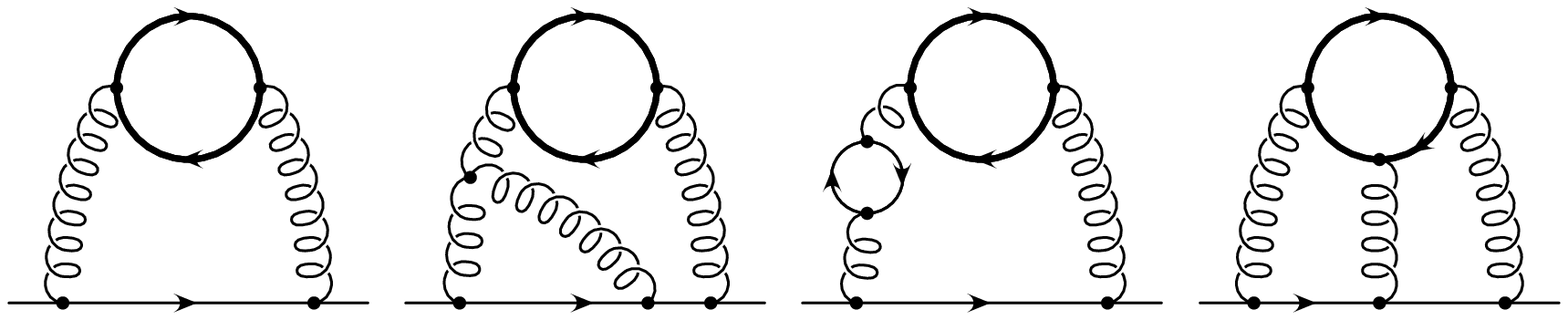,height=2cm,%
bbllx=74pt,bblly=624pt,bburx=569pt,bbury=725pt}}

\medskip

\mbox{\epsfig{file=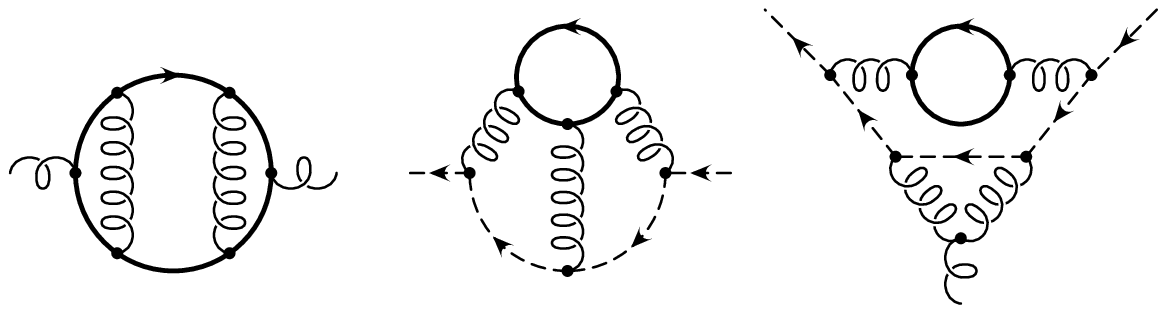,height=2cm,%
bbllx=136pt,bblly=634pt,bburx=470pt,bbury=722pt}}
\caption{Typical Feynman diagrams contributing to $\zeta_2^0$, $\zeta_m^0$,
$\zeta_3^0$, $\tilde\zeta_3^0$, and $\tilde\zeta_1^0$.}
\label{f:one}
\end{figure}

As an example, let us consider the $q$-quark propagator.
Up to terms of ${\cal O}(\mu/m_h)$, we have\cite{alp}
\begin{eqnarray}
\frac{i}{\not\!p\left[1+\Sigma_V^0(p^2)\right]}
&=&
\int dx\,e^{ip\cdot x}\left\langle T\psi_q^0(x)\bar\psi_q^0(0)\right\rangle
\nonumber\\
&=&
\frac{1}{\zeta_2^0}
\int dx\,e^{ip\cdot x}\left\langle T\psi_q^{0\prime}(x)\bar\psi_q^{0\prime}(0)
\right\rangle
=\frac{1}{\zeta_2^0}\,
\frac{i}{\not\!p\left[1+\Sigma_V^{0\prime}(p^2)\right]},
\end{eqnarray}
where the subscript $V$ reminds us that the self-energy of a massless quark
only consists of a vector part.
Note that $\Sigma_V^{0\prime}(p^2)$ only contains light degrees of freedom,
whereas $\Sigma_V^0(p^2)$ also receives virtual contributions from the $h$ 
quark.
As we are interested in the limit $m_h\to\infty$, we may nullify the external
momentum $p$, which entails an enormous technical simplification because then
only tadpole integrals have to be considered.
In dimensional regularization, one also has $\Sigma_V^{0\prime}(0)=0$.
Thus, we obtain 
\begin{equation}
\zeta_2^0=1+\Sigma_V^{0h}(0),
\label{eq:two}
\end{equation}
where the superscript $h$ indicates that only diagrams involving closed
$h$-quark loops need to be computed.
In a similar fashion, one obtains
\begin{eqnarray}
\zeta_m^0&=&\frac{1-\Sigma_S^{0h}(0)}{1+\Sigma_V^{0h}(0)},\qquad
\zeta_3^0=1+\Pi_G^{0h}(0),\qquad
\tilde\zeta_3^0=1+\Pi_c^{0h}(0),
\nonumber\\
\tilde\zeta_1^0&=&1+\Gamma_{G\bar cc}^{0h}(0,0),\qquad
\zeta_g^0=\frac{\tilde\zeta_1^0}{\tilde\zeta_3^0(\zeta_3^0)^{1/2}},
\label{eq:zet}
\end{eqnarray}
where $\Sigma_S$, $\Pi_G$, $\Pi_c$, and $\Gamma_{G\bar cc}$ denote the scalar 
part of the $q$-quark self-energy, the gluon self-energy, the ghost 
self-energy, and the $G\bar cc$ vertex function, respectively.
Typical Feynman diagrams contributing to $\zeta_2^0$, $\zeta_m^0$,
$\zeta_3^0$, $\tilde\zeta_3^0$, and $\tilde\zeta_1^0$ are depicted in 
Fig.~\ref{f:one}.
The full set of diagrams are generated and evaluated with the symbolic
manipulation packages QGRAF\cite{nog} and MATAD,\cite{ste} respectively.
The renormalized counterparts of $\zeta_g^0$ and $\zeta_m^0$,
\begin{equation}
\zeta_g=\frac{Z_g}{Z_g^\prime}\zeta_g^0,\qquad
\zeta_m=\frac{Z_m}{Z_m^\prime}\zeta_m^0,
\end{equation}
are found to be UV finite and $\xi$ independent and to satisfy the appropriate
renormalization group equations, which constitutes a strong test.
The resulting decoupling relations take a particularly simple form if the
matching scale is chosen to be $\mu_h=m_h(\mu_h)$, namely,
\begin{eqnarray}
\frac{a^\prime}{a}&=&\zeta_g^2=1+c_2a^2+c_3a^3,\qquad
\frac{m_q^\prime}{m_q}=\zeta_m=1+d_2a^2+d_3a^3,
\nonumber\\
c_2&=&\frac{11}{72},\qquad
c_3=\frac{564731}{124416}-\frac{82043}{27648}\zeta(3)
-\frac{2633}{31104}n_l,\qquad
d_2=\frac{89}{432},
\nonumber\\
d_3&=&\frac{2951}{2916}-\frac{\ln^42}{54}+\frac{\ln^22}{9}\zeta(2)
-\frac{407}{864}\zeta(3)+\frac{103}{72}\zeta(4)
-\frac{4}{9}\li\left(\frac{1}{2}\right)
\nonumber\\
&&{}+n_l\left(\frac{1327}{11664}-\frac{2}{27}\zeta(3)\right),
\label{eq:coe}
\end{eqnarray}
where $\zeta$ and $\li{}$ are Riemann's zeta function and the dilogarithm, 
respectively.
$c_2$ and $d_2$ were previously calculated.\cite{ber}
Three-loop expressions for $\zeta_2$ and $\zeta_3$, which may be useful for
parton model calculations, are available for the covariant gauge.\cite{dec}

\begin{figure}
\centering
\begin{tabular}{cc}
\mbox{\epsfig{file=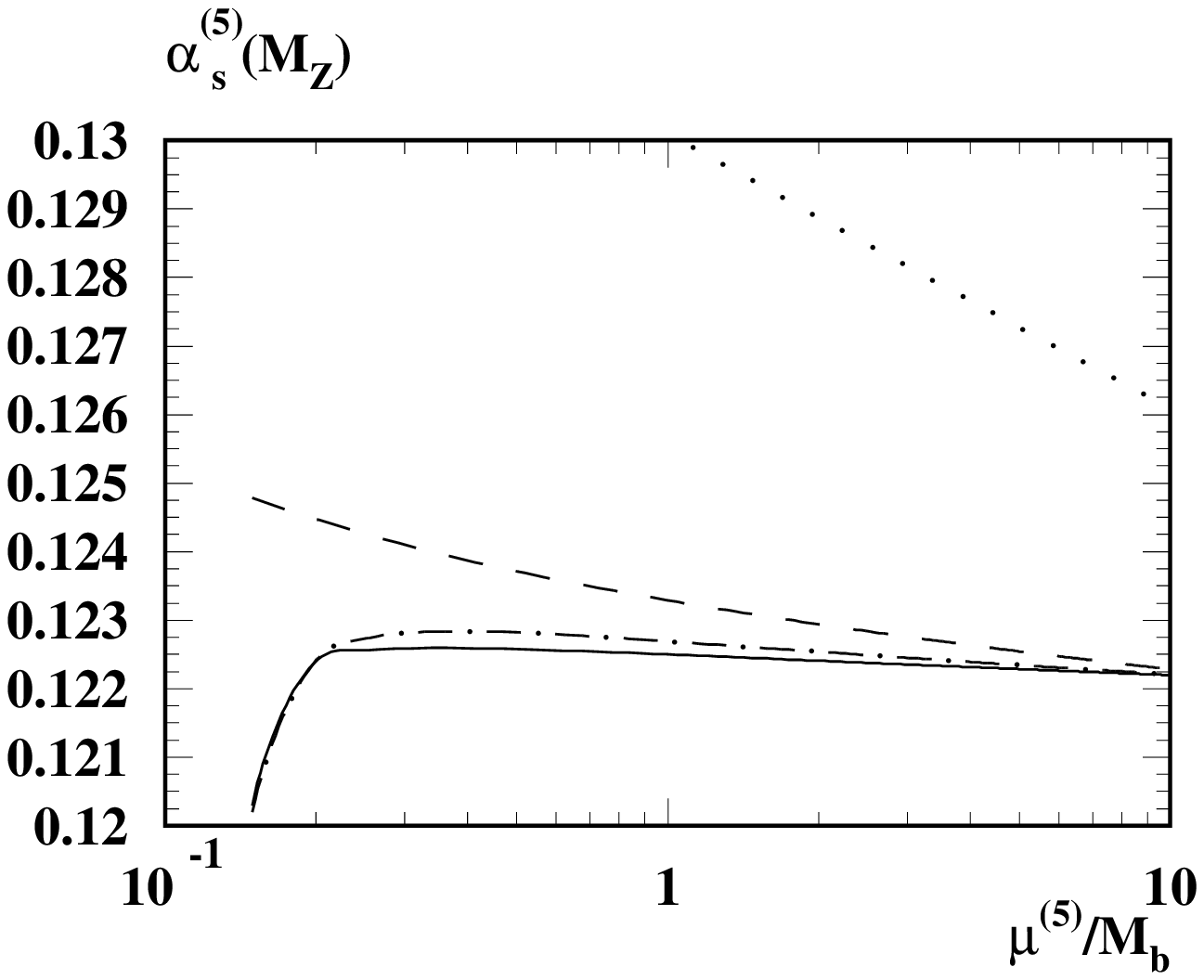,width=5cm,%
bbllx=90pt,bblly=275pt,bburx=463pt,bbury=579pt}}
&
\mbox{\epsfig{file=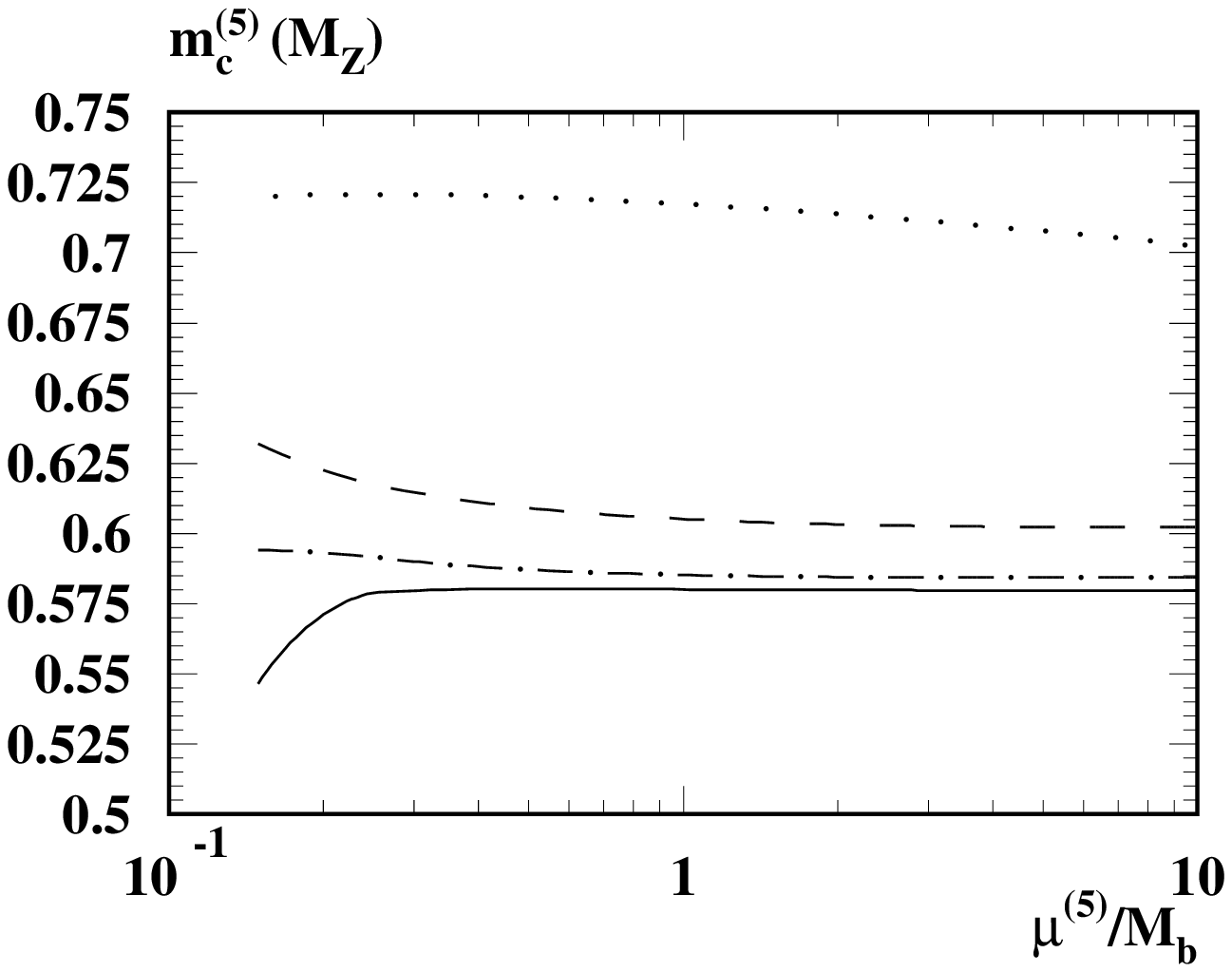,width=5cm,%
bbllx=99pt,bblly=275pt,bburx=463pt,bbury=562pt}}
\end{tabular}
\caption{$\mu^{(5)}$ dependence of (a) $\alpha_s^{(5)}(M_Z)$ calculated from
$\alpha_s^{(4)}(M_\tau)=0.36$ and (b) $m_c^{(5)}(M_Z)$ calculated from 
$\mu_c=m_c^{(4)}(\mu_c)=1.2$~GeV and $\alpha_s^{(5)}(M_Z)=0.118$ with
evolution at one (dotted), two (dashed), three (dot-dashed), and four (solid)
loops and appropriate matching.}
\label{f:two}
\end{figure}

The phenomenological implications of Eqs.~(\ref{eq:rge}) and (\ref{eq:coe})
are illustrated in Fig.~\ref{f:two}.
For consistency, $(n+1)$-loop evolution must be accompanied by $n$-loop 
matching.
Figure~\ref{f:two}(a) shows how $\alpha_s^{(5)}(M_Z)$, consistently evaluated
from $\alpha_s^{(4)}(M_\tau)=0.36$ to a given order, depends on the scale
$\mu^{(5)}$, measured in units of the bottom-quark pole mass $M_b=4.7$~GeV,
where the bottom-quark threshold is crossed.
In Fig.~\ref{f:two}(b), the analogous study is performed for $m_c^{(5)}(M_Z)$
calculated from $\mu_c=m_c^{(4)}(\mu_c)=1.2$~GeV using 
$\alpha_s^{(5)}(M_Z)=0.118$.
As expected, the dependence on the unphysical scale $\mu^{(5)}$ is gradually
getting weaker as we go to higher orders.

\section{Effective Lagrangians and Low-Energy Theorems}

An interesting and perhaps even surprising aspect of $\zeta_g$ and $\zeta_m$
is that they carry the full information about the virtual $h$-quark effects on
the couplings of a CP-even Higgs boson $H$ to gluons and $q$ quarks,
respectively.
To reveal this connection, starting from the bare Yukawa Lagrangian of the
full theory,
\begin{equation}
{\cal L}_{\rm Yuk}=-\frac{H^0}{v^0}\left(\sum_{q=1}^{n_l}m_q^0
\bar\psi_q^0\psi_q^0+m_h^0\bar\psi_h^0\psi_h^0\right),
\end{equation}
where $v$ is the Higgs vacuum expectation value, one integrates out the $h$ 
quark by taking the limit $m_h^0\to\infty$ and so derives the effective
Lagrangian,
\begin{eqnarray}
{\cal L}_{\rm Yuk}^\prime=-\frac{H^0}{v^0}\sum_{i=1}^5C_i^0{\cal O}_i^\prime
=-2^{1/4}G_F^{1/2}H\sum_{i=1}^5C_i[{\cal O}_i],
\label{eq:yuk}
\end{eqnarray}
which is spanned by a natural basis of composite scalar operators with mass 
dimension four.\cite{ina}
The operators,
\begin{eqnarray}
{\cal O}_1^\prime&=&\left(G_{\mu\nu}^{0\prime,a}\right)^2,\qquad
{\cal O}_2^\prime=\sum_{q=1}^{n_l}m_q^{0\prime}\bar\psi_q^{0\prime}
\psi_q^{0\prime},\qquad
{\cal O}_3^\prime=
\sum_{q=1}^{n_l}\bar\psi_q^{0\prime}
\left(i\not\!\!D^{0\prime}-m_q^{0\prime}\right)\psi_q^{0\prime},
\nonumber\\
{\cal O}_4^\prime&=&G_\nu^{0\prime,a}
\left(\nabla^{ab}_\mu G^{0\prime,b\mu\nu}
+g_s^{0\prime}\sum_{q=1}^{n_l}\bar\psi_q^{0\prime}T^a\gamma^\nu
\psi_q^{0\prime}\right)
-\partial_\mu \bar{c}^{0\prime,a}\partial^\mu c^{0\prime,a},
\nonumber\\
{\cal O}_5^\prime&=&
(\nabla^{ab}_\mu\partial^\mu\bar{c}^{0\prime,b})c^{0\prime,a},
\end{eqnarray}
are only constructed from light degrees of freedom, while all residual
dependence on the $h$ quark resides in the Wilson coefficients $C_i^0$.

The derivation of $C_i^0$ proceeds similarly to Eq.~(\ref{eq:zet}).
Considering appropriate one-particle-irreducible Green functions which
contain a zero-momentum insertion of ${\cal O}_h=m_h^0\bar\psi_h^0\psi_h^0$ in
the limit $m_h^0\to\infty$, one finds\cite{dec}
\begin{eqnarray}
&&\zeta_3^0(-4C_1^0+2C_4^0)=-\frac{1}{2}\partial_h^0\Pi_G^{0h}(0),\
\zeta_m^0\zeta_2^0(C_2^0-C_3^0)
=1-\Sigma_S^{0h}(0)-\frac{1}{2}\partial_h^0\Sigma_S^{0h}(0),
\nonumber\\
&&\zeta_2^0C_3^0=-\frac{1}{2}\partial_h^0\Sigma_V^{0h}(0),\
\tilde\zeta_3^0(C_4^0+C_5^0)=\frac{1}{2}\partial_h^0\Pi_c^{0h}(0),\
\tilde\zeta_1^0C_5^0=\frac{1}{2}\partial_h^0\Gamma_{G\bar cc}^{0h}(0,0),
\label{eq:wil}
\end{eqnarray}
with $\partial_h^0=(m_h^{02}\partial/\partial m_h^{02})$, which may be solved
for $C_i^0$.
Only ${\cal O}_1^\prime$ and ${\cal O}_2^\prime$ contribute to physical
observables.
They mix under renormalization as\cite{ina}
\begin{equation}
\left[{\cal O}_1^\prime\right]=
\left[1+2\left(\frac{\alpha_s^\prime\partial}{\partial\alpha_s^\prime}
\ln Z_g^\prime\right)\right]{\cal O}_1^\prime
-4\left(\frac{\alpha_s^\prime\partial}{\partial\alpha_s^\prime}
\ln Z_m^\prime\right){\cal O}_2^\prime,\qquad
\left[{\cal O}_2^\prime\right]={\cal O}_2^\prime,\end{equation}
where the brackets denote the renormalized counterparts.
$C_1$ and $C_2$ are accordingly determined from the second equation in
Eq.~(\ref{eq:yuk}).
They are diagrammatically calculated through three loops.\cite{dec}
Inserting Eqs.~(\ref{eq:two}) and (\ref{eq:zet}) into Eqs.~(\ref{eq:wil}), one 
obtains the low-energy theorems\cite{dec}
\begin{equation}
C_1=-\frac{1}{2}\,\frac{\partial\ln\zeta_g^2}{\partial\ln m_h^2},\qquad
C_2=1+2\frac{\partial\ln\zeta_m}{\partial\ln m_h^2},
\label{eq:let}
\end{equation}
which are valid to all orders in $\alpha_s$.
Fully exploiting the present knowledge of Eq.~(\ref{eq:rge}),\cite{ver} one
may construct the four-loop terms of $\zeta_g$ and $\zeta_m$ involving
$\ln m_h^2$ and so obtain $C_1$ and $C_2$ from Eq.~(\ref{eq:let}) to one order 
beyond the diagrammatic calculation.
The expansions in $a=\alpha_s^{(n_f)}(\mu_h)/\pi$ read\cite{dec}
\begin{eqnarray}
C_1&=&-\frac{a}{12}[1+2.7500\,a+(9.7951-0.6979\,n_l)a^2
\nonumber\\
&&{}+(49.1827-7.7743\,n_l-0.2207\,n_l^2)a^3],
\nonumber\\
C_2&=&1+0.2778a^2+(2.2434+0.2454\,n_l)a^3
\nonumber\\
&&{}+(2.1800+0.3096\,n_l-0.0100\,n_l^2)a^4.
\end{eqnarray}

Having established ${\cal L}_{\rm yuk}^\prime$, we are able to make 
higher-order predictions for the QCD interactions of a light $H$ boson by
just computing massless diagrams.
For instance, the $H\to gg$ partial decay width at three loops is found to
be\cite{hgg}
\begin{equation}
\Gamma(H\to gg)=\frac{G_FM_H^3}{36\pi\sqrt2}a^{\prime2}
\left[1+17.917\,a^\prime
+a^{\prime2}\left(156.808-5.708\,\ln\frac{m_t^2}{M_H^2}\right)\right],
\label{eq:hgg}
\end{equation}
where $a^\prime=\alpha_s^{(5)}(M_H)/\pi$.
The three-loop ${\cal O}(\alpha_s^2G_Fm_t^2)$ corrections to
$\Gamma(H\to q\bar q)$, with $q=u,d,s,c,b$, may also be obtained from
Eq.~(\ref{eq:yuk}).\cite{hbb}
Analogously, the QCD interactions of a CP-odd Higgs boson $A$ may be described
by an effective Lagrangian involving composite pseudoscalar operators with
mass dimension four.\cite{agg}
The resulting counterpart of Eq.~(\ref{eq:hgg}) is found to be\cite{agg}
\begin{equation}
\Gamma(A\to gg)=\frac{G_FM_A^3}{16\pi\sqrt2}a^{\prime2}
\left[1+18.417\,a^\prime
+a^{\prime2}\left(171.544-5\ln\frac{m_t^2}{M_A^2}\right)\right],
\label{eq:agg}
\end{equation}
where $a^\prime=\alpha_s^{(5)}(M_A)/\pi$.
As a by-product of this analysis,\cite{agg} the Adler-Bardeen
nonrenormalization theorem,\cite{adl} which states that the anomaly of the
axial-vector current is not renormalized in QCD, is verified through three
loops by an explicit diagrammatic calculation.

\section{Comparison with Scale Optimization Procedures}

It is interesting to compare the exact values of the ${\cal O}(\alpha_s^2)$
corrections in Eqs.~(\ref{eq:hgg}) and (\ref{eq:agg}) with the estimates one
may derive from the knowledge of the ${\cal O}(\alpha_s)$ correction through
the application of well-known scale optimization procedures, based on 
Grunberg's concept of fastest apparent convergence (FAC), Stevenson's
principle of minimal sensitivity (PMS), and the proposal by Brodsky, Lepage,
and Mackenzie (BLM) to resum the leading light-quark contribution to the
renormalization of the strong coupling constant.\cite{gru}
The resulting estimates are listed in Table~\ref{t:one}.
We observe that the sign and the order of magnitude is correctly predicted in
all cases.

\begin{table}[ht]
\begin{center}
\caption{Scale optimization estimates for the ${\cal O}(\alpha_s^2)$
coefficients in Eqs.~(\ref{eq:hgg}) and (\ref{eq:agg}).}
\label{t:one}
\medskip
\begin{tabular}{|c|c|c|c|} \hline\hline
& FAC & PMS & BLM \\
\hline
$H\to gg$ & 263.3 & 263.9 & 242.5 \\
\hline
$A\to gg$ & 277.6 & 278.1 & 252.5 \\
\hline\hline
\end{tabular}
\end{center}
\end{table}

\section{Summary}

A consistent $\overline{\mbox{MS}}$ description of $\alpha_s(\mu)$ and
$m_q(\mu)$ with $\mu$ evolution through four loops and threshold matching
through three loops is now available.
Effective Lagrangians and low-energy theorems are useful tools to treat the
hadronic decays of light CP-even and CP-odd Higgs bosons through three loops.
The sign and the order of magnitude of the resulting three-loop corrections
are correctly predicted by scale optimization procedures.

\section*{Acknowledgments}
The author is grateful to W.A. Bardeen, K.G. Chetyrkin, and M. Steinhauser for
their collaboration and to the organizers of the IVth International Symposium
on Radiative Corrections (RADCOR~98) for their excellent work.

\end{document}